\documentstyle[prb,aps,multicol,epsfig]{revtex}

\begin{document}

\draft

\title{Corrections to Scaling in the Hydrodynamic Properties
       of Dilute Polymer Solutions}

\author{
Burkhard D\"unweg\footnote{corresponding author, Electronic
    Mail: \texttt{duenweg@mpip-mainz.mpg.de}}, Dirk Reith,
    Martin Steinhauser, and Kurt Kremer\\
Max--Planck--Institut f\"ur Polymerforschung \\
Ackermannweg 10, D--55128 Mainz, Germany }

\date{\today}
\maketitle

\widetext

\begin{abstract}
  We discuss the hydrodynamic radius $R_H$ of polymer chains in good solvent,
  and show that the leading order correction to the asymptotic law $R_H
  \propto N^\nu$ ($N$ degree of polymerization, $\nu \approx 0.59$) is an
  ``analytic'' term of order $N^{-(1 - \nu)}$, which is directly related to
  the discretization of the chain into a finite number of beads. This result
  is further corroborated by exact calculations for Gaussian chains, and
  extensive numerical simulations of different models of good--solvent chains,
  where we find a value of $1.591 \pm 0.007$ for the asymptotic universal
  ratio $R_G / R_H$, $R_G$ being the chain's gyration radius. For $\Theta$
  chains the data apparently extrapolate to $R_G / R_H \approx 1.44$, which is
  different from the Gaussian value $1.5045$, but in accordance with previous
  simulations. We also show that the experimentally observed deviations of the
  initial decay rate in dynamic light scattering from the asymptotic
  Benmouna--Akcasu value can partly be understood by similar arguments.
\end{abstract}

\begin{multicols}{2}
\narrowtext

\section{Introduction and Summary}
\label{sec:intro}

It is well--known that the average size $R$ of an isolated flexible uncharged
polymer chain in good solvent is asymptotically proportional to $N^\nu$, where
$N$ is the degree of polymerization, and $\nu \approx 0.5877$
\cite{limadrassokal}. This law holds for any measure of the chain size, the
most popular of which are the mean square end--to--end distance,
\begin{equation} \label{eq:endenddistance}
\left< R_E^2 \right> = \left< r_{1 N}^2 \right> ,
\end{equation}
the mean square radius of gyration,
\begin{equation} \label{eq:gyrationradius}
\left< R_G^2 \right> = \frac{1}{2 N^2} \sum_{ij} \left< r_{i j}^2 \right> ,
\end{equation}
and the hydrodynamic radius
\begin{equation} \label{eq:hydrodynradius}
\left< \frac{1}{R_H} \right> = \frac{1}{N^2} \sum_{i \ne j}
\left< \frac{1}{r_{i j}} \right> .
\end{equation}
In these equations, we have assumed that the chain is composed of $N$ monomers
(i.~e. $N - 1$ bonds) at positions $\vec r_i$, $i = 1, \ldots, N$, and $r_{i
  j} = \left\vert \vec r_i - \vec r_j \right\vert$. Experimentally, the
gyration radius is determined from small--angle scattering experiments; for
small wave numbers $q$ the single--chain static structure factor behaves like
\cite{doiedw}
\begin{eqnarray} \label{eq:structurefactor}
S(q) & = & \frac{1}{N} \sum_{i j}
\left< \exp \left[ i \vec q \cdot \vec r_{i j} \right] \right>
\nonumber \\
& = & N \left[ 1 - \frac{q^2}{3} \left< R_G^2 \right> + O(q^4) \right] . 
\end{eqnarray}
Conversely, the hydrodynamic radius is determined via small--angle dynamic
light scattering experiments, where the dynamic structure factor
\begin{equation} \label{eq:dynstrucfac}
S(q,t) = \frac{1}{N}
\sum_{i j} \left< \exp\left[ i \vec q \cdot 
\left( \vec r_i(t) - \vec r_j (0) \right) \right] \right> 
\end{equation}
is measured. In the small $q$ limit, it decays like
\begin{equation} \label{eq:decaydynstruc1}
\frac{S(q,t)}{S(q,0)} = \exp \left( - D q^2 t \right),
\end{equation}
where $D$ is the chain diffusion constant, which, within the framework of
Kirkwood--Zimm theory \cite{doiedw} is related to $R_H$ via
\begin{equation} \label{eq:kirkwoodformula}
D = \frac{D_0}{N} + \frac{k_B T}{6 \pi \eta} \left< \frac{1}{R_H} \right> ,
\end{equation}
where $D_0$ is the monomer diffusion constant (usually this contribution
is neglected), $k_B$ is Boltzmann's constant, $T$ the temperature and
$\eta$ the solvent viscosity. In principle, it is possible to obtain
$R_H$, as a static quantity, also from purely static scattering, as
is seen from the relation
\begin{equation} \label{eq:rhfromsofq}
\left< \frac{1}{R_H} \right> = \frac{2}{\pi N} \int_0^\infty dq
\left( S(q) - 1 \right),
\end{equation}
but for technical reasons, this has so far not been applied in experiments.

When analyzing data for the chain size, one has to take into account that the
law $R \propto N^\nu$ holds only in the asymptotic limit $N \to \infty$, while
for finite chain lengths deviations occur. This is particularly important for
computer simulations, where data with high statistical accuracy can be
obtained. For this reason, corrections to scaling have been worked out in
great detail, and exploited in high--resolution numerical studies, for the
end--to--end distance and the gyration radius, where the relation
\begin{equation} \label{eq:knowncorrtoscaling}
\left< R_E^2 \right> = A N^{2 \nu}
\left( 1 + \frac{B}{N^\Delta} + \ldots \right)
\end{equation}
(and analogously for $R_G$) holds \cite{sokal}. Here $A$ and $B$ are
non--universal amplitudes, while $\Delta$ is a universal
correction--to--scaling exponent, whose value is difficult to determine beyond
the accuracy $\Delta \approx 0.5$ ($\Delta \approx 0.56$ according to Ref.
\onlinecite{limadrassokal}, $\Delta \approx 0.43$ according to Ref.
\onlinecite{besold}). The omitted terms are further powers $N^{-\Delta - 1}$,
$N^{-\Delta - 2}$, \dots, as well as $N^{-\Delta_2}$, $N^{-\Delta_2 - 1}$,
\ldots (i.~e. there are further larger correction--to--scaling exponents),
plus so--called ``analytic'' terms $N^{-1}$, $N^{-2}$, \ldots \cite{sokal}.
The important point to notice is that the ``analytic'' corrections will arise
even for a Gaussian chain, and are due to the fact that the chain consists of
a finite number of beads. This will be demonstrated explicitly in Sec.
\ref{sec:hydradgaussian}. Conversely, the ``non--analytic'' corrections are
due to the fact that, in the language of renormalization group (RG) theory,
the chain's Hamiltonian is not identical to the fixed point Hamiltonian. The
exponent $\Delta$ is related to the largest sub--leading eigenvalue of the RG
transformation at the fixed point. In first order $\epsilon$ expansion its
value is \cite{schaefer} $\Delta = \nu \omega$ with $1 / \nu = 2 - \epsilon /
4 + O(\epsilon^2)$ and $\omega = \epsilon + O (\epsilon^2)$, where $\epsilon =
4 - d$ and $d$ is the spatial dimension. Higher--order calculations
\cite{zinnjustin} have resulted in $\Delta = \nu \omega = 0.588 \times 0.82 =
0.482$. We adopt here the convention (which we view as quite natural) to
distinguish the terms by their different origins, and call correction terms
``analytic'' corrections if they are present even in the Gaussian limit, while
we call ``non--analytic'' corrections those terms which occur exclusively for
excluded--volume chains. As we will see in Sec. \ref{sec:hydradgoodsolv},
``analytic'' corrections defined in this way do not necessarily imply integer
powers of $N$.

As for the hydrodynamic radius of self--avoiding walks (SAWs), there is no
high--resolution numerical study available, and corrections to scaling have
not yet been dealt with systematically. This is somewhat unfortunate, since
the corrections are unusually large for $R_H$, and of experimental relevance.
For a good solvent chain, one expects again $N^{-\Delta}$ etc. terms, plus
``analytic'' corrections. It is the main purpose of the present paper to show
that the leading--order term of these latter corrections is now given by
\begin{equation} \label{eq:analcorrhydrad}
\left< \frac{1}{R_H} \right> = \frac{A}{N^\nu}
       \left( 1 - \frac{B}{N^{1 - \nu}} + \ldots \right) ,
\end{equation}
where $B$ is usually positive. We will show in Secs. \ref{sec:hydradgoodsolv}
and \ref{sec:altderiv} that this form is a straightforward consequence of
discretizing the chain into beads. As $1 - \nu \approx 0.41$, this will, for
long chains, ultimately dominate over the $N^{-\Delta}$ term, where the
exponent is (according to Refs. \onlinecite{limadrassokal},
\onlinecite{besold} and \onlinecite{zinnjustin}) slightly larger.
Nevertheless, the exponents are so close that in most cases one will observe
contributions from both terms. On the other hand, it is a well--known
empirical fact that for many experimental systems, as well as for most
computer models, the corrections to scaling of the gyration radius are quite
weak, such that the $N^{-\Delta}$ term should have a rather small amplitude.
One could therefore expect that the corresponding amplitude of the
$N^{-\Delta}$ contribution in $R_H$ is also quite small. Then the most likely
candidate for explaining the experimental and numerical observation that $R_H$
is usually subject to very large corrections to scaling \cite{schaefer} would
actually be the ``analytic'' $N^{- (1 - \nu)}$ term.

For a {\em Gaussian} chain we are able to solve the problem exactly, see Sec.
\ref{sec:hydradgaussian}:
\begin{equation} \label{eq:gaussmainz}
\left< \frac{b}{R_H} \right> = 
\frac{8}{3} \left( \frac{6}{\pi} \right)^{1/2} N^{-1/2}
       \left( 1 - B N^{-1/2} + \ldots \right) ,
\end{equation}
with $B = - (3/4) \zeta(1/2) \approx 1.095266$ (here $\zeta$ is Riemann's zeta
function), and $b$ denoting the root mean square bond length. As $\nu = 1/2$
for a Gaussian chain, this form is consistent with Eq.
\ref{eq:analcorrhydrad}.

The difficulties in observing the asymptotic $N^\nu$ scaling of $R_H$ have a
long history. Adam and Delsanti \cite{adamdelsanti1,adamdelsanti2} performed
dynamic light scattering experiments and found an effective power law $R_H
\propto N^{0.55}$. This is quite typical, and has been found in many other
experiments, too \cite{nemoto,venkataswamy,simone}, although an exponent of
$0.61$ \cite{tsunashima1} has been reported as well. A reduction of the
effective exponent is indeed expected, as seen from Eq.
\ref{eq:analcorrhydrad}, and is also observed in Brownian Dynamics simulations
\cite{depablo}. As a caveat, note that a scattering experiment does not
measure $R_H$, but rather the diffusion constant $D$. This quantity has an
additional $D_0 / N$ contribution (Eq. \ref{eq:kirkwoodformula}), which is of
the same order as the leading correction of Eq. \ref{eq:analcorrhydrad}.
Therefore, the corrections in $D$ are weaker than those in $R_H$.
Nevertheless, the $D_0 / N$ term is typically not large enough to fully
compensate the corrections in $R_H$. This is easily seen for the Gaussian case
from Eqs. \ref{eq:kirkwoodformula} and \ref{eq:gaussmainz}: The monomer
diffusion constant $D_0$ can be written as $D_0 = k_B T / (6 \pi \eta a)$,
which defines a monomer Stokes radius $a$. Thus
\begin{eqnarray}
\frac{D}{D_0} & = & N^{-1} \\
& + & \frac{a}{b} \left( 3.6853 N^{-1/2}
      - 4.0364 N^{-1} + \ldots \right) . \nonumber
\end{eqnarray}
Since $a$ should be of the order of the bond length $b$, one sees that
a large $N^{-1}$ contribution remains.

A first attempt to explain the experimental observation is due to Weill and
des Cloizeaux \cite{weilldescloizeaux}. They conjectured that $\nu_{eff} =
0.55$ is due to non--perfect solvent quality, and a crossover between good
solvent behavior ($\nu \approx 0.6$, large length scales) and $\Theta$ solvent
($\nu = 0.5$, small length scales). In particular, they pointed out that the
averaging over $1 / r$ assigns a very large statistical weight to the small
distances. Although this latter argument is true, and generally accepted as
the basic origin for the slow convergence of $R_H$, the explanation in terms
of solvent quality has turned out to be incorrect. In Ref.
\onlinecite{schaeferbaumgaertner} it was demonstrated that $R_H$ should {\em
  not} be much more susceptible to solvent quality effects than $R_E$ or $R_G$
--- the enhanced sensitivity of $R_H$ to the small distances is balanced by
the fact that $R_E$ and $R_G$ are more sensitive to the decreased swelling of
the chain near its ends: A SAW is {\em inhomogeneous}, i.~e. $\left< r_{ij}^2
\right>$ depends on the position of the $ij$ bond on the chain, and is
systematically larger in the interior, as has been shown both numerically
\cite{schaeferbaumgaertner,kkdipl} and analytically
\cite{schaefer,schaeferbaumgaertner}.

Furthermore, Sch\"afer and Baumg\"artner \cite{schaeferbaumgaertner} performed
a detailed RG calculation and predicted in one--loop order for the universal
amplitude ratio
\begin{eqnarray} \label{eq:rgrhoneloop}
\rho_\infty & = & \lim_{N \to \infty} \rho(N)
= \lim_{N \to \infty} \frac{R_G(N)}{R_H(N)} \nonumber \\
& \approx & 1.06 \times \frac{8}{3 \sqrt{\pi}}
\approx 1.06 \times 1.5045 \approx 1.595 ;
\end{eqnarray}
here $8 / (3 \sqrt{\pi})$ is the exact random walk (RW) value. Other RG
studies resulted in $\rho_\infty = 1.562$ \cite{oonokohmoto} (this value was
later revised to 1.51, see Ref. \onlinecite{tsunanew}) and $\rho_\infty =
1.62$ \cite{douglasfreed}, while a semi--empirical relation based on fitting
the distribution function of internal distances to light scattering data
yields $\rho_\infty = 1.5955$ \cite{tsunashima1,tsunanew}. A value of
$\rho_\infty \approx 1.6$ was also found in Brownian Dynamics simulations
\cite{depablo}.  {From} Eq. \ref{eq:analcorrhydrad} it is clear that $\rho$
should be subject to an $N^{-(1 - \nu)}$ correction for finite chain length;
nevertheless, experiments have so far not reported a systematic dependence on
molecular weight. Typically, values around $\rho \approx 1.5$
\cite{simone,tsunashima1}, or $\rho \approx 1.6$ / $\rho \approx 1.3$ for
different solvents \cite{venkataswamy} are found in the good--solvent regime.
In view of the inaccuracies of the experiments ($R_G$ typically has an error
of $5 \%$ \cite{venkataswamy}) the inability to observe a systematic behavior
in $N$ is not very surprising.

In order to contribute to the resolution of these questions, we have performed
computer simulations of very different models of polymer chains, both for SAWs
and for $\Theta$ chains, and calculated $R_G$ and $R_H$, as outlined in Sec.
\ref{sec:numres}. To provide a complete and well--converged data set
represents the second main goal of our paper. To our knowledge, our results
are the most accurate data obtained on $R_H$ so far. Concerning $R_G$,
however, our data are less accurate than those of Li {\em et al.}
\cite{limadrassokal}, which we still view as the most precise numerical study
on the SAW problem so far. Therefore we have taken their values for the
exponents $\nu$ and $\Delta$ for our fits. We find $\rho_\infty = 1.591
\pm 0.007$ for good--solvent chains, in very good agreement with Ref.
\onlinecite{schaeferbaumgaertner}. (Note that our error estimate is probably
overly optimistic, since it only includes statistical errors and completely
neglects systematic errors.)

Theoretical and numerical investigations on corrections to scaling in $R_H$
have first focused on the RW case. The work by Guttman {\em et al.}
\cite{guttman1,guttman2,akcasnew} showed by analytical calculation that a
Gaussian chain should obey Eq. \ref{eq:gaussmainz}. The prefactor of the
correction was first \cite{guttman1} determined only approximately, $B \approx
1.125$, while later \cite{akcasnew} it was given exactly in terms of an
integral. Furthermore, Monte Carlo (MC) simulations of lattice chains at the
$\Theta$ point revealed that in this case the ratio $R_G / R_H$ apparently
does {\em not} converge to its Gaussian value $1.5045$, but rather to roughly
$1.4$. Our simulations (see Sec. \ref{sec:numres}) find a similar behavior
($\rho_\infty \approx 1.44$). We believe that this can be explained
qualitatively from RG arguments \cite{schaefer} as follows: The asymptotic
behavior is expected to be governed by the Gaussian fixed point, and thus
$\rho_\infty$, as a universal amplitude ratio, is expected to assume the
Gaussian value. However, the numerical extrapolation will only produce this
value if all relevant correction terms, i.~e. the analytic $N^{-1/2}$ term,
plus the non--analytic corrections, are consistently taken into account.
Neither the data analysis by Guttman {\em et al.}  \cite{guttman1,guttman2},
nor ours, fulfill this requirement, as both just fit to Eq.
\ref{eq:analcorrhydrad} with $\nu = 1/2$, and thus are expected to produce
substantial systematic errors in $\rho_\infty$. To do this in a better way is
practically impossible, since (i) our $\Theta$ data have insufficient
statistical accuracy to allow for additional fit parameters, (ii) the precise
form of the correction terms is unknown for $R_H$ (in contrast to $R_E$ and
$R_G$, for which the leading--order terms have been calculated by tricritical
field theory \cite{hager}, with the interesting feature that they are
universal), and (iii) the non--analytic corrections vary extremely slowly
(logarithmically) with $N$, such that either one would need unrealistically
long chains to ensure dominance of the leading orders, or an expansion up to
unrealistically high order. These problems have been elucidated in quite some
detail for $R_E$ and $R_G$ \cite{hager}, explaining previous difficulties in
the interpretation of highly accurate MC data on $\Theta$ chains
\cite{grassberger}. In this context, it should be mentioned that experiments
\cite{schmidtburchard,tsunashima2} typically find a value of $\rho = 1.3$,
i.~e. a similar reduction as in the good solvent case.

Later, MC data were taken of excluded--volume (EV) chains with SAW statistics.
Sch\"afer and Baumg\"artner \cite{schaeferbaumgaertner} used chains of up to
$161$ monomers, with an EV strength particularly close to the SAW fixed point,
such that poor--solvent effects can be ruled out. The inhomogeneous swelling
was demonstrated, and the $R_H$ data were fitted with Eq.
\ref{eq:analcorrhydrad}. This was done with an empirical
correction--to--scaling exponent of $1/2$ instead of $1 - \nu$. The same
evidence was shown in the simulation data by Batoulis and Kremer
\cite{batouliskremer} of chains of length of up to $N \approx 400$. Ladd and
Frenkel \cite{laddfrenkel} simulated chains of length of up to $N = 1025$ and
were able to describe their $R_H$ data via Eq. \ref{eq:analcorrhydrad}, with
$A = 3.84$ and $B = 1.06$, but without detailed justification of their use of
the correct $1 - \nu$ exponent. Sch\"afer and Baumg\"artner
\cite{schaeferbaumgaertner} concluded from both their analytical studies and
their simulation data that not the solvent quality, but rather the chain's
microstructure is responsible for the slow convergence. Our reasoning (Secs.
\ref{sec:hydradgaussian}--\ref{sec:altderiv}), which is similar to the one by
Guttman {\em et al.} \cite{guttman1,guttman2,akcasnew}, exactly supports this
picture: The corrections are due to the fact that the chain is discretized
into beads, or, in other words, to the fact that there is a lower length scale
cutoff for the frictional properties. However, the notion of ``stiffness'',
which is often used in this context \cite{schaeferbaumgaertner}, is, in our
view, somewhat misleading: As outlined in Sec. \ref{sec:stiffness}, we expect
a large local chain stiffness to {\em decrease} the correction until it
ultimately even changes its sign. The same conclusion has been found by Akcasu
and Guttman \cite{akcasnew} for stiff chains without excluded volume.

In the context of dynamic light scattering of dilute polymer solutions there
is yet another unresolved puzzle. As Akcasu {\em et al.} have shown
\cite{akcasugurol}, the initial decay rate of the dynamic structure factor,
\begin{equation} \label{eq:omegaofqdef}
\Omega (q) = \left. \frac{d}{dt} \frac{S(q,t)}{S(q,0)} \right\vert_{t = 0} ,
\end{equation}
can be written as
\begin{equation} \label{eq:omegaofqakcasu}
\Omega (q) = \frac{\sum_{ij}
\left< \vec q \cdot \tensor D_{ij} \cdot \vec q
\exp( i \vec q \cdot \vec r_{ij} ) \right> }{
\sum_{ij} \left<
\exp( i \vec q \cdot \vec r_{ij} ) \right> } ,
\end{equation}
where $\tensor D_{ij}$ is the diffusion tensor. Equation
\ref{eq:omegaofqakcasu} is a rigorous result, the only assumption being that
the chain dynamics can be described by Kirkwood's diffusion equation
\cite{doiedw}. Usually, $\tensor D_{ij}$ is taken as the Oseen tensor,
\begin{equation} \label{eq:oseendef}
\tensor D_{ij} = D_0 \delta_{ij} \tensor 1 + (1 - \delta_{ij})
\frac{k_B T}{8 \pi \eta r_{ij}}
( \tensor 1 + \hat r_{ij} \otimes \hat r_{ij} ) ,
\end{equation}
where $\hat r_{ij} \otimes \hat r_{ij}$ denotes the tensor product of the unit
vector in $\vec r_{ij}$ direction with itself. In this case, Eq.
\ref{eq:omegaofqakcasu} is just the $q > 0$ generalization of Eq.
\ref{eq:kirkwoodformula}. It can then be shown \cite{doiedw,benmounaakcasu}
that for $q$ in the scaling regime $R_G^{-1} \ll q \ll b^{-1}$
($b$ denoting the bond length), or, strictly spoken, in the
limit $q b \to 0$, $q R_G \to \infty$, the relation
\begin{equation} \label{eq:plateau}
\Omega (q) = C \frac{k_B T}{\eta} q^3
\end{equation}
holds, where the numerical constant $C$ only depends on chain statistics: $C =
0.0625$ for RW statistics ($\nu = 1/2$) and $C = 0.0788$ for SAWs ($\nu =
0.6$). This has been tested by light scattering experiments both for good
solvents \cite{nemoto,tsunashima1,hanakcasu,bhatt1,bhatt2} and for $\Theta$
solvents \cite{tsunashima2,hanakcasu}. In both cases the relation is verified
with reasonable accuracy, but with a prefactor $C$ which is systematically
smaller than the theoretical prediction. The reasons for this shift are not
clear; an attempt by a generalized theory which introduces draining
\cite{shiwa} so far had only limited success \cite{tsunashima3}. In Sec.
\ref{sec:initdecay} we show that the deviation can partly be explained by the
fact that in reality neither $q b = 0$ nor $q R_G = \infty$ holds. Taking
these nonidealities crudely into account, we find a shift in the same
direction, which is however smaller than the experimental one. Nevertheless,
we believe that this third main result is of direct relevance for the analysis
of experimental data. There are also some indications from Molecular Dynamics
simulations \cite{bdphd} that the description in terms of the Kirkwood theory
is insufficient on these length and time scales.

\section{Analytical Theory}
\label{sec:theory}

\subsection{Hydrodynamic Radius of a Gaussian Chain}
\label{sec:hydradgaussian}

For a Gaussian chain with root mean square bond length $b$, we have
\begin{equation}
\left< r_{ij}^2 \right> = b^2 \left\vert i - j \right\vert
\end{equation}
and
\begin{equation}
\left< r_{ij}^{-1} \right> = 6^{1/2} \pi^{-1/2}
b^{-1} \left\vert i - j \right\vert^{-1/2} ,
\end{equation}
and hence
\begin{equation}
\left< R_E^2 \right> = b^2 \left(N - 1 \right) 
= b^2 N \left(1 - \frac{1}{N} \right)
\end{equation}
and
\begin{eqnarray}
\left< R_G^2 \right> & = & \frac{b^2}{N^2} \sum_{i < j} (j - i)
= \frac{b^2}{N^2} \sum_{n = 1}^{N - 1} n (N - n) \nonumber \\
& = & \frac{1}{6} b^2 N \left( 1 - \frac{1}{N^2} \right) ,
\end{eqnarray}
where we have used elementary summation formulae. For the hydrodynamic
radius, we find analogously
\begin{equation}
\left< R_H^{-1} \right> = \sqrt{ \frac{6}{\pi} }
\frac{2}{b N^2} \sum_{n = 1}^{N - 1} n^{-1/2} (N - n) .
\end{equation}
According to the Euler--Maclaurin formula (see Appendix \ref{sec:euler}), Eq.
\ref{eq:eulermaclaurinpower}, the sums can be expanded as
\begin{eqnarray}
\sum_{n = 1}^{N - 1} n^{-1/2} & = & 
                     2 N^{1/2} - \frac{1}{2} N^{-1/2} 
                     + \zeta\left(\frac{1}{2}\right) 
                     + O(N^{-3/2}) , \nonumber \\
\sum_{n = 1}^{N - 1} n^{+1/2} & = &
                     \frac{2}{3} N^{3/2} - \frac{1}{2} N^{1/2}
                     \nonumber \\
                     & + & \zeta\left(- \frac{1}{2}\right)
                     + O(N^{-1/2}) .
\end{eqnarray}
Hence,
\begin{equation}
\sum_{n = 1}^{N - 1} n^{-1/2} (N - n) =
\frac{4}{3} N^{3/2} + N \zeta \left( \frac{1}{2} \right) + O(N^0)
\end{equation}
and
\begin{eqnarray}
\left< R_H^{-1} \right> & = & \sqrt{ \frac{6}{\pi} }
\frac{8}{3 b} N^{-1/2} \times \nonumber \\
& & \left( 1 + \frac{3}{4} \zeta \left( \frac{1}{2} \right) N^{-1/2}
    + O(N^{-3/2}) \right) ,
\end{eqnarray}
which is the result anticipated in Eq. \ref{eq:gaussmainz}.

\subsection{Hydrodynamic Radius of a Good Solvent Chain}
\label{sec:hydradgoodsolv}

For a linear SAW, the main difficulty is the fact that, unlike for a RW,
$\left< r_{ij}^2 \right>$ and $\left< r_{ij}^{-1} \right>$ do {\em not}
\cite{schaefer,schaeferbaumgaertner,kkdipl} just depend on $\left\vert i - j
\right\vert$, but rather on the positions relative to the ends as well. In
order to obtain the leading--order analytic corrections due to discretization,
we can restrict the discussion to the leading--order scale--invariant
behavior, i.~e. we can assume that the SAW is strictly scale invariant with
the exponent $\nu$, with no non--analytic corrections. If we would include the
latter, they would just generate further additive terms in our expressions. In
what follows, we therefore omit them, for the sake of simplified notation, but
keep in mind that they have to be added at the end in order to obtain the full
expressions. We thus assume the relations
\begin{eqnarray}
\phi_G \left(\lambda x, \lambda y \right) & = & \lambda^{2 \nu}
\phi_G \left(        x,         y \right) ,  \\
\phi_H \left(\lambda x, \lambda y \right) & = & \lambda^{- \nu}
\phi_H \left(        x,         y \right) ,
\end{eqnarray}
where we have introduced the notation
\begin{eqnarray}
\phi_G \left(i, j \right) & = & \left< r_{ij}^2 \right> , \\
\phi_H \left(i, j \right) & = & \left< r_{ij}^{-1} \right> .
\end{eqnarray}
The definitions of $R_G$ and $R_H$ lead us to study the sum
\begin{equation}
\sigma (N) = \sum_{n = 2}^N \sum_{m = 1}^{n - 1} \phi( m, n )
\end{equation}
for $\phi = \phi_G$ and $\phi = \phi_H$, respectively, by means of the
Euler--Maclaurin expansion of Appendix \ref{sec:euler}. Treating the inner sum
first, we find
\begin{equation} \label{eq:eulersaw1}
\sum_{m = 1}^{n - 1} \phi(m, n) = \mbox{\rm const.} + \varphi(n)
\end{equation}
with the formal expansion
\begin{equation} \label{eq:eulersaw2}
\varphi(n) = \int_1^n dx \phi(x, n) 
+ \frac{1}{12} \left. \frac{d}{dx} \phi (x, n) \right\vert_{x = n}
+ \ldots ,
\end{equation}
since $\phi(n,n)$ vanishes. Note also that the constant in Eq.
\ref{eq:eulersaw1} does not depend on $n$, hence
\begin{eqnarray}
\sigma (N) & = & (N - 1) \mbox{\rm const.} + 
\sum_{n = 2}^N \varphi(n) \nonumber \\ & = &
(N - 1) \mbox{\rm const.} +
\int_2^{N + 1} dy \varphi (y) \nonumber \\
& - & \frac{1}{2} \varphi(N + 1)
+ \mbox{\rm const.} + \ldots .
\end{eqnarray}
Inserting Eq. \ref{eq:eulersaw2}, we find

\begin{figure}[t]
\psfig{figure=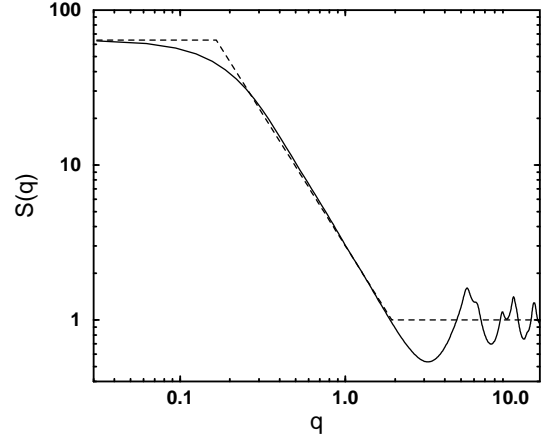,width=7cm}
\caption{$S(q)$ as defined in Eq. \ref{eq:structurefactor}, for a
  chain of length $N = 64$, for model C (see Sec. \ref{sec:numres}),
  demonstrating the $q^{-1 / \nu}$ decay, followed by oscillations around
  unity. The dashed line is the simplified structure factor of Eq.
  \ref{eq:simplemodelstructurefactor}.}
\label{fig:sofq}
\end{figure}

\begin{eqnarray}
\sigma (N)
     & = & \int_2^{N + 1} dy \int_1^y dx \phi(x, y) \nonumber \\
     & + & \int_2^{N + 1} dy 
           \frac{1}{12} \left. \frac{d}{dx} \phi (x, y) \right\vert_{x = y}
           \nonumber \\
     & - & \frac{1}{2} \int_1^{N + 1} dx \phi(x, N + 1) \nonumber \\
     & + & N \mbox{\rm const.} \nonumber \\
     & + & \mbox{\rm const.}  + \ldots .
\end{eqnarray}
After transformation to the reduced variables $u = x / N$ and $v = y / N$, and
exploiting the scaling behavior of $\phi$, it is possible to determine the
order of each term. For the gyration radius, we find
$O(N^{2 + 2\nu})$, 
$O(N^{2 \nu})$,
$O(N^{1 + 2\nu})$, 
$O(N^{1})$, 
$O(N^{0})$, 
respectively, for the five terms in order of their appearance. Conversely, for
$\left< R_H^{-1} \right>$ the different scaling behavior of $\phi$ implies
$O(N^{2 - \nu})$, 
$O(N^{- \nu})$, 
$O(N^{1 - \nu})$,
$O(N^{1})$, 
$O(N^{0})$ 
for the corresponding orders. For $\left< R_G^2 \right>$ the leading order is
$O(N^{2 + 2\nu})$, while the next sub--leading order is $O(N^{1 + 2\nu})$,
resulting in a leading order correction of $O(N^{-1})$. For $\left< R_H^{-1}
\right>$, the leading order is $O(N^{2 - \nu})$, followed by the $O(N^{1})$
term. Thus the correction to scaling for $R_G$ has order $O(N^{-(1 - \nu)})$.
This proves Eq. \ref{eq:analcorrhydrad}. Of course, this consideration does
not prove that the amplitude $B$ in Eq. \ref{eq:analcorrhydrad} is positive;
however, this is expected from the result for Gaussian chains.

\subsection{Alternative Derivation}
\label{sec:altderiv}

Equation \ref{eq:analcorrhydrad} can also be derived in a more heuristic way,
starting from Eq. \ref{eq:rhfromsofq}. Figure \ref{fig:sofq} shows the typical
behavior of $S(q)$: For wave numbers $q$ with $R_G^{-1} \ll q \ll b^{-1}$ the
structure factor exhibits a power--law decay $q^{-1/\nu}$ which indicates the
chain's fractal geometry, while for larger $q$ it oscillates around unity. We
therefore can introduce a cutoff wavenumber $q_0$ from which on there is no
further contribution to the integral, i.~e. $q_0$ is the smallest of all the
$\hat q$'s with the property $\int_{\hat q}^\infty dq (S(q) - 1) = 0$. Hence,
\begin{equation}
\frac{1}{R_H} = 
\frac{2}{\pi N} \int_0^{q_0} dq S(q) - \frac{2 q_0}{\pi N} .
\end{equation}
It is physically clear that for a flexible chain $q_0$ must be roughly $(2
\pi) / b$, apart from a numerical prefactor of order unity. Moreover, the
fractal $q^{-1/\nu}$ decay of $S(q)$ roughly extends up to $q_0$, at which
point $S(q) \approx 1$ is reached. We now introduce a modified structure
factor $\tilde S (q)$, which is identical to $S(q)$ up to $q = q_0$, but
extends the $q^{-1 / \nu}$ decay up to $q = \infty$. In this latter regime,
we have
\begin{equation}
\tilde S(q) = \alpha \left( \frac{q}{q_0} \right)^{-1/\nu} ,
\end{equation}
where $\alpha$ is a numerical prefactor of order unity. Therefore
we can write
\begin{equation}
\frac{1}{R_H} = 
\frac{2}{\pi N} \int_0^{\infty} dq \tilde S(q) 
- \frac{2}{\pi N} \int_{q_0}^\infty dq \tilde S(q)
- \frac{2 q_0}{\pi N} .
\end{equation}
Evaluating the second integral, and writing $\tilde S (q)$ in scaling form,
\begin{equation} \label{eq:strufacscalingform}
\tilde S(q) = N s(q R_G) 
\end{equation}
(here we have again assumed strict scale invariance, i.~e. absence of
non--analytic corrections to scaling, for the same reason as outlined
at the beginning of the previous subsection), one finds
\begin{equation}
\frac{R_G}{R_H} = \frac{2}{\pi} \int_0^{\infty} dx s(x)
- \frac{2}{\pi} \left( \alpha \frac{\nu}{1 - \nu} + 1 \right)
\frac{q_0 R_G}{N} ,
\end{equation}
i.~e. again a negative correction of order $O(N^{-(1 - \nu)})$.

\subsection{Effect of Chain Stiffness}
\label{sec:stiffness}

The advantage of the approach of the previous subsection is that it can be
easily generalized to study the influence of local structure, since it is
well--known how this is reflected in $S(q)$. For a locally stiff chain with a
persistence length large compared to the bond length $b$, one expects that
$q_0$ is roughly unchanged with respect to the flexible case. However, the
$q^{-1 / \nu}$ decay does no longer extend down to $q \approx q_0$, but only
to $q \approx q_1$, where $q_1$ is a crossover wave number, whose inverse is a
typical length scale below which stiffness effects are important. With $\tilde
S (q)$ being again the continuation of the $q^{-1 / \nu}$ decay up to $q =
\infty$, we have
\begin{eqnarray} \label{eq:splittheintegral}
\frac{1}{R_H} & = & \frac{2}{\pi N} \int_0^\infty     dq \tilde S (q)
                  - \frac{2}{\pi N} \int_{q_1}^\infty dq \tilde S (q)
                    \nonumber \\
              & + & \frac{2}{\pi N} \int_{q_1}^{q_0}  dq S (q)
                  - \frac{2 q_0}{\pi N} .
\end{eqnarray}
We now assume
\begin{equation}
\tilde S(q) = \alpha \frac{q_0}{q_1}
\left( \frac{q}{q_1} \right)^{-1 / \nu}
\end{equation}
for $q > q_1$, and
\begin{equation}
S(q) = \beta \left( \frac{q}{q_0} \right)^{-1}
\end{equation}
for $q_1 < q < q_0$. Here, $\alpha$ and $\beta$ denote prefactors of order
unity, and the $q^{-1}$ decay results from the local stretching. Evaluating
the integrals, and using Eq. \ref{eq:strufacscalingform}, one thus finds
\begin{eqnarray}
\frac{R_G}{R_H} & = & \frac{2}{\pi} \int_0^{\infty} dx s(x) \\
&& - \frac{2}{\pi} \left( \alpha \frac{\nu}{1 - \nu} + 1 
- \beta \ln \frac{q_0}{q_1} \right) \frac{q_0 R_G}{N} . \nonumber
\end{eqnarray}
In order to compare with the flexible case, we still have to take into account
that stiffness tends to increase the gyration radius, by roughly a factor of
$(q_0 / q_1)^{1 - \nu}$:
\begin{eqnarray}
\frac{R_G}{R_H} & = & \frac{2}{\pi} \int_0^{\infty} dx s(x) \\
&& - \frac{2}{\pi} \left( \frac{q_0}{q_1} \right)^{1 - \nu}
\left( \alpha \frac{\nu}{1 - \nu} + 1 
- \beta \ln \frac{q_0}{q_1} \right) \frac{q_0 R_G^{(0)}}{N} ,
\nonumber
\end{eqnarray}
where $R_G^{(0)}$ denotes the gyration radius in the flexible case.

The prefactor of the correction term hence depends on the stiffness parameter
$q_0 / q_1$ in a non--trivial way; for small $q_0 / q_1$ both an increase and
a decrease are possible, depending on the parameters. For sufficiently large
stiffness one always obtains a decrease of the correction, and ultimately even
a change of its sign.

\subsection{Initial Decay Rate}
\label{sec:initdecay}

In this subsection, we are concerned with the initial decay rate $\Omega (q)$,
see Eq. \ref{eq:omegaofqakcasu}. Splitting the sum in the numerator into
diagonal and off--diagonal terms, one finds
\begin{eqnarray}
\Omega (q) & = & \frac{D_0 q^2}{S(q)} \\
& + & \frac{1}{N S(q)}
\sum_{i \ne j}
\left< \vec q \cdot \tensor D_{ij} \cdot \vec q
\exp( i \vec q \cdot \vec r_{ij} ) \right> . \nonumber
\end{eqnarray}
Following Refs. \onlinecite{doiedw,hammouda}, we use the Fourier
representation of the Oseen tensor for the off--diagonal elements,
\begin{equation}
\tensor D_{ij} = \frac{k_B T}{\eta}
\frac{1}{(2 \pi)^3} \int d^3 k
\frac{\tensor 1 - \hat k \otimes \hat k}{k^2}
\exp(i \vec k \cdot \vec r_{ij}) ,
\end{equation}
to find
\begin{eqnarray}
\Omega (q) & = & \frac{D_0 q^2}{S(q)} + \frac{1}{S(q)} \frac{k_B T}{\eta}
\frac{1}{(2 \pi)^3} \int d^3 k \nonumber \\
& & \times \frac{q^2 - (\hat k \cdot \vec q)^2}{k^2}
\left( S(\vec k + \vec q) - 1 \right) .
\end{eqnarray}
We now focus attention on the dimensionless quantity
\begin{eqnarray}
C(q) & = & \frac{\eta}{q^3 k_B T} \Omega(q) \nonumber \\
     & = & \frac{1}{6 \pi q a S(q)} +
           \frac{1}{S(q)} \frac{1}{(2 \pi)^3} \int d^3 k \nonumber \\
     &&    \times  \frac{1 - (\hat k \cdot \hat q)^2}{q k^2}
           \left( S(\vec k + \vec q) - 1 \right)  ,
\end{eqnarray}
where we again have expressed the monomer diffusion constant $D_0$ in terms of
a Stokes radius $a$. After transforming to the dimensionless integration
variable
\begin{equation}
\vec x = \frac{ \vec k + \vec q }{q}
\end{equation}
and performing the angular integration, one has \cite{doiedw,hammouda}
\begin{eqnarray}
C(q) & = & \frac{1}{6 \pi q a S(q)} \\
&& + \frac{1}{S(q)} \frac{1}{(2 \pi)^2}
\int_0^\infty dx f(x) \left( S(q x) - 1 \right) 
\nonumber
\end{eqnarray}
with
\begin{equation}
f(x) = x^2 \left( \frac{1 + x^2}{2 x} 
\ln \left\vert \frac{1 + x}{1 - x} \right\vert - 1 \right) .  
\end{equation}
This function can be expanded as
\begin{equation}
f(x) = \sum_{n = 0}^\infty \left( 
\frac{1}{2 n + 1} + \frac{1}{2 n + 3} \right) x^{2 n + 4}
\end{equation}
for $x < 1$, and
\begin{equation}
f(x) = \sum_{n = 0}^\infty \left( 
\frac{1}{2 n + 1} + \frac{1}{2 n + 3} \right) x^{- 2 n}
\end{equation}
for $x > 1$.

In order to make further progress, we have to specify the structure factor
$S(q)$. This shall be done by the most simplistic model which takes into
account both finite bead size and finite chain length (see also Fig.
\ref{fig:sofq}):
\begin{equation} \label{eq:simplemodelstructurefactor}
S(q) = \left\{
\begin{array}{l l}
N                                            & \hspace{0.5cm}
                                             q < \frac{2 \pi}{a} N^{- \nu} \\
\left( \frac{q a}{2 \pi} \right)^{-1 / \nu}  & \hspace{0.5cm}
                                             \frac{2 \pi}{a} N^{- \nu}
                                             < q < \frac{2 \pi}{a} \\
1                                            & \hspace{0.5cm}
                                             q > \frac{2 \pi}{a}
\end{array}
\right. .
\end{equation}
We now calculate $C(q)$ in the scaling regime $R_G^{-1} \ll q \ll a^{-1}$.
Defining the $x$ values where $S(q x)$ changes its behavior as
\begin{eqnarray}
x_1 & = & \frac{2 \pi}{q a N^\nu} \ll 1 , \\
x_2 & = & \frac{2 \pi}{q a} \gg 1 ,
\end{eqnarray}
we can write $(q a)^{-1} = x_2 / (2 \pi)$, $S(q)^{-1} = x_2^{-1 / \nu}$,
$N / S(q) = x_1^{- 1 / \nu}$; hence
\begin{eqnarray}
C(q) & = & \frac{1}{12 \pi^2} x_2^{1 - 1 / \nu} \nonumber \\
     & + & \frac{1}{(2 \pi)^2} x_1^{- 1 / \nu} \int_0^{x_1} dx f(x)
                                              \nonumber \\
     & + & \frac{1}{(2 \pi)^2} \int_{x_1}^{x_2} dx f(x) x^{-1 / \nu}
                                              \nonumber \\
     & - & \frac{1}{(2 \pi)^2} x_2^{-1 / \nu} \int_0^{x_2} dx f(x) . 
\end{eqnarray}
Since $x_1 \ll 1$ and $x_2 \gg 1$, we can write
\begin{eqnarray}
\int_0^{x_1} dx f(x) & \approx & \frac{4}{15} x_1^5, \\
\int_0^{x_2} dx f(x) & \approx & \frac{4}{3} x_2,
\end{eqnarray}
where we have taken just the leading--order terms of the expansions of $f$;
this results in
\begin{eqnarray}
C(q) & \approx & \frac{1}{12 \pi^2} x_2^{1 - 1 / \nu} \nonumber \\
     & + & \frac{1}{ 15 \pi^2} x_1^{5 - 1 / \nu}
                                              \nonumber \\
     & + & \frac{1}{(2 \pi)^2} \int_{x_1}^{x_2} dx f(x) x^{-1 / \nu}
                                              \nonumber \\
     & - & \frac{1}{3 \pi^2} x_2^{1 -1 / \nu} . 
\end{eqnarray}
In the asymptotic limit $q R_G \to \infty$, i.~e. $x_1 \to 0$, and
$q a \to 0$, i.~e. $x_2 \to \infty$, this obviously converges to
the asymptotic value
\begin{eqnarray}
C_{as} & = & \frac{1}{(2 \pi)^2} \int_0^\infty dx f(x) x^{- 1 / \nu} 
\nonumber \\
& = & \left\{
\begin{array}{l l l}
1 / 16             & =       0.0625 & \hspace{0.5cm} \nu = 1 / 2 \\
\sqrt{3} / (7 \pi) & \approx 0.0788 & \hspace{0.5cm} \nu = 3 / 5
\end{array}
\right. .
\end{eqnarray}
Focusing now on the correction, i.~e. $\Delta C(q) = C(q) - C_{as}$,
we find
\begin{eqnarray}
\Delta C(q) & \approx & \frac{1}{12 \pi^2} x_2^{1 - 1 / \nu} \nonumber \\
     & + & \frac{1}{ 15 \pi^2} x_1^{5 - 1 / \nu}
                                              \nonumber \\
     & - & \frac{1}{(2 \pi)^2} \int_{0}^{x_1} dx f(x) x^{-1 / \nu}
                                              \nonumber \\
     & - & \frac{1}{(2 \pi)^2} \int_{x_2}^{\infty} dx f(x) x^{-1 / \nu}
                                              \nonumber \\
     & - & \frac{1}{3 \pi^2} x_2^{1 - 1 / \nu} ;
\end{eqnarray}
taking again the leading--order terms for the remaining integrals results in
\begin{eqnarray} \label{eq:shiftresult}
\Delta C(q) & \approx & - \frac{1}{12 \pi^2} \frac{3 + \nu}{1 - \nu}
                          x_2^{1 - 1 / \nu} \nonumber \\
            &&          - \frac{1}{15 \pi^2} \frac{1}{5 \nu - 1} 
                          x_1^{5 - 1/ \nu} .
\end{eqnarray}
One thus sees that {\em both} finite chain length {\em and} finite bead size
have the tendency to decrease $C$, as observed in the experiments. The latter
effect is clearly more important, as $x_1$ enters only via a relatively high
power. Further insight is gained by numerical evaluation of the shift for
reasonable parameter values.

Tsunashima {\em et al.} \cite{tsunashima1} performed their experiments with
polyisoprene chains of size $R_G = 210 nm$. Typical scattering wavenumbers in
their plateau regime were given by $q R_G = 4 \ldots 8$; the experimental
observation in this regime was $C \approx 0.06$, i.~e. a shift of $\Delta C
\approx - 2 \times 10^{-2}$. In what follows, we consider the value $q R_G
\approx 6$. Thus
\begin{equation}
x_2 = \frac{2 \pi}{q R_G} \frac{R_G}{a} \approx 500,
\end{equation}
where we have estimated the monomer size $a$ as $0.45 nm$ \cite{tsunaprivate}.
Inserting this into Eq. \ref{eq:shiftresult}, we find for the $x_2$
contribution a value of $\Delta C \approx - 1 \times 10^{- 3}$, i.~e. one
order of magnitude smaller than the experimental value.

It is not completely clear if a more thorough treatment of the integral would
fully account for the deviation; our guess is that it would probably not.
Molecular Dynamics data \cite{bdphd} seem to rather indicate that for typical
systems (i.~e. on not yet asymptotic length scales) the coupling between
polymer and solvent is more complex than the simple Kirkwood description.
Nevertheless, we consider our result as important for the interpretation of
experimental data: There is obviously a substantial contribution to $\Delta C$
which stems from the finite bead size, and which is only weakly
$q$--dependent. A plateau--like shape of $C(q)$ alone apparently does not
guarantee asymptotic behavior. Clearly more work has to be done to fully
resolve the puzzle, but we believe our considerations show that theories which
neglect the influence of finite bead size (and, to a lesser degree, of finite
chain length) are simply not accurate enough to describe experimental data
even of rather long chains.

\section{Numerical Results}
\label{sec:numres}

In our numerical studies, we have used three different polymer chain models,
which we will denote as model A, B, and C.

{\em Model A} is a bead--spring model in the continuum. $N$ monomers
are connected via an anharmonic (``finitely extensible nonlinear elastic'')
spring potential,
\begin{equation} \label{eq:fene}
  U_{FENE} = \left\{
    \begin{array}{ll}
      -\frac{1}{2} k R_{0}^2 \ln
      \left[ 1-\left(\displaystyle\frac{r}{R_0}\right)^2 \right]
             & \hspace{0.5cm} r < R_0\\
      \infty & \hspace{0.5cm} r \geq R_0
    \end{array}
\right. ,
\end{equation}
where we use the standard parameters \cite{kremer} $k = 30$, $R_0 = 1.5$ in
dimensionless units. Between all monomers there is an additional
non--bonded potential
\begin{equation} \label{eq:ljcos}
  U_{LJcos} = \left\{%
    \begin{array}{l}
      4 \left[ \left( \displaystyle\frac{1}{r} \right)^{12}
        -\left( \displaystyle\frac{1}{r} \right)^6 
        + \frac{1}{4} \right] - \lambda ,
        \hspace{0.2cm} r \le 2^{1/6} \\
        \frac{1}{2} \lambda \left[ \cos ( \alpha r^2 +\beta ) - 1 \right] ,
        \hspace{0.2cm} 2^{1/6} \le r \le 1.5 \\
       \\
      0 , \hspace{0.2cm} r \ge 1.5 ,
    \end{array}\right. 
\end{equation}
where $\alpha$ and $\beta$ are determined as the solutions of the linear set
of equations
\begin{eqnarray}
  \label{eq:lineq}
  2^{1/3}\alpha  + \beta &=& \pi\\
  2.25   \alpha  + \beta &=& 2\pi ,
\end{eqnarray}
i.~e. $\alpha = 3.1730728678$ and $\beta = -0.85622864544$. This potential has
originally been constructed to simulate amphiphilic systems \cite{thoso}. The
parameter $\lambda$ serves to control the strength of the attractive
interaction and is varied instead of the temperature, which is fixed at $k_B T
= 1$. For sufficiently strong $\lambda$, the chain assumes a collapsed state,
while $\lambda = 0$ corresponds to good solvent. We used a combination of
stochastic dynamics \cite{kremer} and the pivot algorithm \cite{sokal}.
Applying standard methods \cite{kremer} on data of chains of length of up to
$N = 2000$, we located the $\Theta$ point at $\lambda = 0.65 \pm 0.02$.
In the good solvent limit, and at $\lambda = 0.65$, we also ran an $N = 5000$
chain.

{\em Model B} is a mesoscopic model for an aqueous solution of the sodium salt
of poly (acrylic acid) (PAA), whose input parameters have been derived from an
extensive atomistic simulation of an aqueous PAA solution ($T=333.15$ K and
$p=1$ atm) in the highly diluted regime, such that the ion concentration
(number of charges on the chain, plus counterions) is $0.4$ mol/l
\cite{biermann01s}. From this simulation, structural averages like the
distributions of bond angles or radial distribution functions between monomers
were extracted. We mapped this system to the mesoscale by replacing one
repeating unit (i.~e.\ one monomer) by one bead. As center of the
coarse--grained (CG) beads, the monomer center of mass (excluding the sodium
ion) was chosen. Bonded as well as non--bonded terms were parameterized by
systematically varying the interactions until the structure of the atomistic
model was reproduced \cite{reith00s}. This also allowed us to neglect all
explicit water molecules and sodium ions (necessarily present in the parent
atomistic simulation) in subsequent CG simulations. Their effect on the PAA
chain conformation is, however, implicitly present in the model. This means
that a system of roughly $10^{4}$ atoms could be reduced to a system which
consists of only $23$ ``super atoms''. As in model A, we used both stochastic

\begin{figure}
\psfig{figure=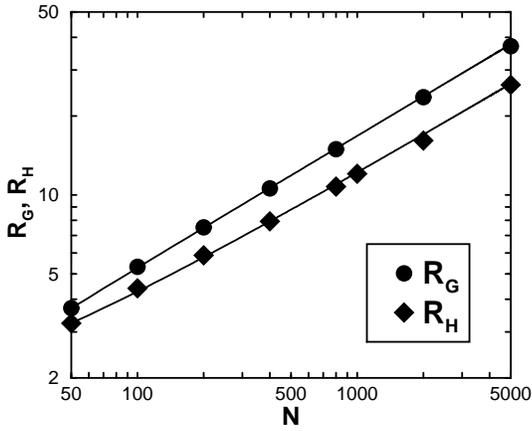,width=7cm}
\caption{$R_G = \left< R_G^2 \right>^{1/2}$ and
  $R_H = \left< R_H^{-1} \right>^{-1}$ for model A at $\Theta$
  condition $\lambda = 0.65$. Error bars are smaller than symbol
  sizes. Also shown are the fits as discussed in the text.}
\label{fig:martintheta}
\end{figure}

\begin{figure}
\psfig{figure=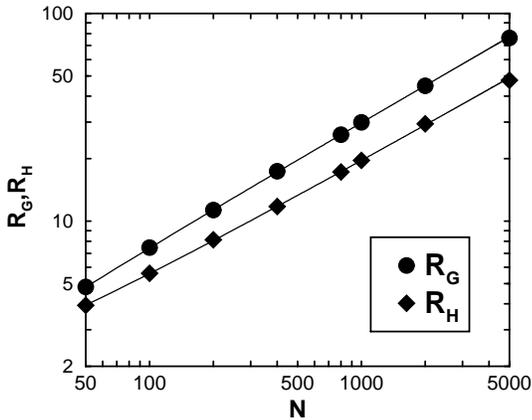,width=7cm}
\caption{Model A: Same as Fig. \ref{fig:martintheta}, but
  for good solvent condition $\lambda = 0$.}
\label{fig:martingood}
\end{figure}

dynamics and pivot Monte Carlo moves. The final force field was utilized to
calculate $R_G$, $R_H$ and other static properties like the structure factor
for PAA strands of length 8 to 3155 repeating units \cite{simone}. The
numerical results agree well with light scattering data on dilute PAA
solutions with corresponding mean molar weights. In particular, the
hydrodynamic radii of six different PAA--salt samples with molecular weights
in the range from $18100$ to $296600$ g/mol were measured. For four samples,
the molar masses $M_{W}$ and the radii of gyration $R_G$ were measured as
well. The PAA samples were of polydispersity $D_P$ between $1.5$ and $1.8$ and
diluted in aqueous NaCl--containing solution ($0.1-1$ mol/l) \cite{simone}.

Finally, {\em model C} is the SAW on the face--centered cubic lattice, which
we prefer over the simple cubic for reasons of increased local flexibility,
which in turn means proximity to the SAW fixed point. Units of length are
defined in such a way that the bond length is $\sqrt{2}$. The chains of length
$N = 64, 128, \ldots 32768$ were generated by using a dimerization procedure
\cite{sokal}. Up to $N = 8192$ the statistical sample always consisted of $M =
1024$ chains, while 

\begin{figure}
\psfig{figure=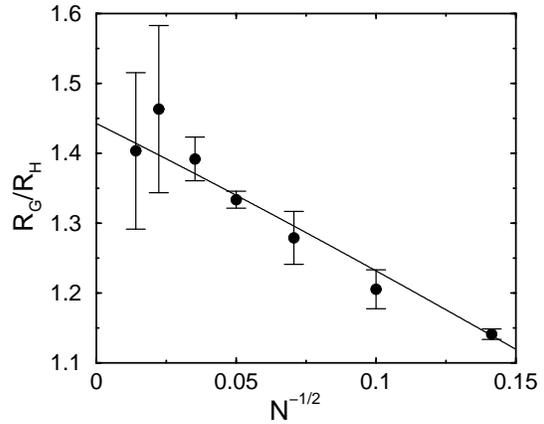,width=7cm}
\caption{Model A: $R_G / R_H$ as a function of the
   scaling variable $N^{-1/2}$, at $\Theta$
   condition $\lambda = 0.65$. The line results
   from the combined fits of $R_G$ and $R_H$.}
\label{fig:ratiomartintheta}
\end{figure}

\begin{figure}
\psfig{figure=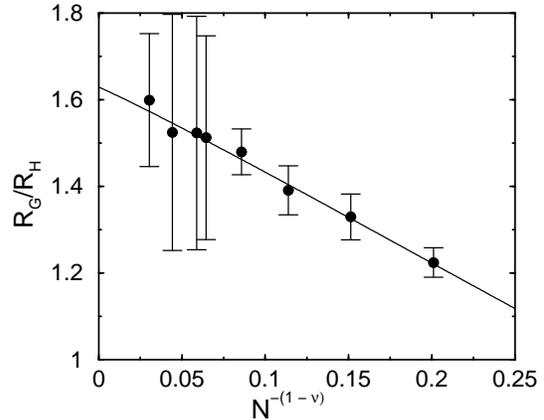,width=7cm}
\caption{Model A: $R_G / R_H$ as a function of the
   scaling variable $N^{-(1 - \nu)}$, at good solvent
   condition $\lambda = 0$. The line results
   from the combined fits of $R_G$ and $R_H$.}
\label{fig:ratiomartingood}
\end{figure}

$M = 1085$ for $N = 16284$ and $M = 296$ for $N = 32768$.

In what follows, we outline our $R_G$ and $R_H$ data for these three models.
Figures \ref{fig:martintheta} and \ref{fig:martingood} summarize our results
for model A at $\Theta$ condition $\lambda = 0.65$ (Fig.
\ref{fig:martintheta}) and at good solvent condition $\lambda = 0$ (Fig.
\ref{fig:martingood}), respectively. For the $\Theta$ chains, we obtained very
good fits with the functions $\left< R_G^2 \right> = 0.2834 N - 0.53$ and
$\left< R_H^{-1} \right> = 2.710 N^{-1/2} - 3.74 N^{-1}$, while for the good
solvent data the analogous fits are $\left< R_G^2 \right> = 0.2706 N^{1.1754}
- 0.32 N^{0.62}$ and $\left< R_H^{-1} \right> = 3.131 N^{-0.5877} - 3.04
N^{-1}$. These fit functions are also shown in Figs. \ref{fig:martintheta} and
\ref{fig:martingood}. The ratio $\rho = R_G / R_H$, as it results from these
data, is shown in the subsequent Figs. \ref{fig:ratiomartintheta} and
\ref{fig:ratiomartingood} for $\Theta$ and good solvent conditions,
respectively. It should be noted that the numerical resolution (for each of
our models) is clearly by far not competitive with the study by Li {\em et
  al.}  \cite{limadrassokal}. For this reason, we did not attempt to determine
the exponents from our data, but rather used the values for $\nu$ and $\Delta$
from Ref. \onlinecite{limadrassokal}. We did not include an $N^{-\Delta}$ term
in the fit for $R_H$ 

\begin{figure}
\psfig{figure=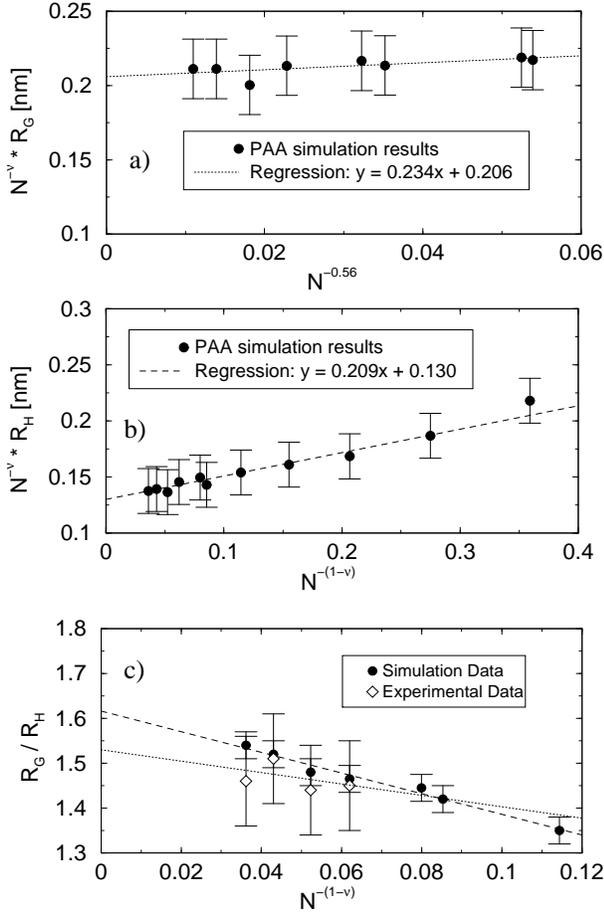,width=8cm}
\caption{Model B: Scaling behavior of poly (acrylic acid) as
  measured by light scattering experiments and computer simulations with a
  coarse--grained model:
  (a) Radius of gyration $R_G$, (b) hydrodynamic radius $R_H$,
  (c) dimensionless ratio $R_G/R_H$. }
\label{fig:RGRH_dirk}
\end{figure}

in the SAW case, although such a term is expected to be
present. The reason is that our model A data are too inaccurate to allow for
such a three--parameter fit in a stable way. Similarly, we ignored the
non--analytic corrections to scaling in the $\Theta$ case, for essentially the
same reason, as has been discussed in some more detail in Sec.
\ref{sec:intro}. Taking the statistical inaccuracies of the data, and of the
resulting fit parameters into account, we obtain for the asymptotic amplitude
ratio $\rho = R_G / R_H$ the values $\rho = 1.44 \pm 0.01$ at the $\Theta$
point, and $\rho = 1.63 \pm 0.01$ in the excluded--volume case. The actual
error in $\rho$ is expected to be significantly larger, since neither the
uncertainties in the exponents and in the location of the $\Theta$ point, nor
systematic errors due to higher--order corrections to scaling have been taken
into account. This is particularly apparent in the $\Theta$ case, where one
expects in the asymptotic long--chain limit rather the Gaussian value
$1.5045$, but also obvious in the SAW case, where the results on the longer
chains of model C yield a considerably smaller value for $\rho$.

The most interesting aspect of model B is that it closely

\begin{figure}
\psfig{figure=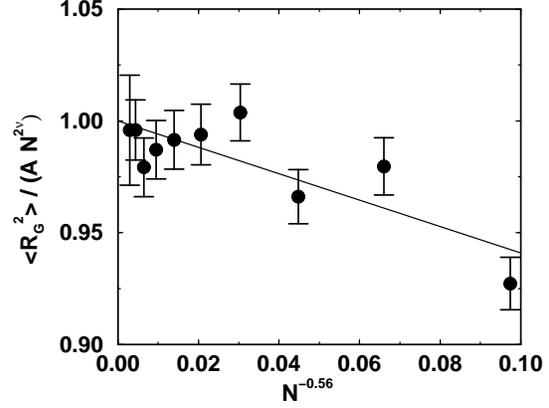,width=7cm}
\caption{Model C: $\left< R_G^2 \right> / (A N^{2 \nu})$
  as a function of $N^{-\Delta}$, where we use $\Delta = 0.56$, and $A
  = 0.3341$ from the (also shown) fit $\left< R_G^2 \right> = A
  N^{1.1754} + B N^{0.62}$.}
\label{fig:bdrg2corscal}
\end{figure}

resembles a real
system, and a quantitative comparison with experiments is possible
\cite{simone}. In Fig. \ref{fig:RGRH_dirk} we show simulation results for
$R_G$, $R_H$, and their ratio. The data are taken as published in Ref.
\onlinecite{simone}. For the ratio, experimental results are also included.
The scaling $N^{-(1 - \nu)}$, and the extrapolation to $\rho = 1.61 \pm 0.02$
is nicely borne out by the simulation data. The experiments are too inaccurate
to demonstrate a clear systematic trend. In spite of this, an extrapolation
yields $\rho \approx 1.5-1.6$, which means that the theoretical calculations
are supported by data of a real chemical system.

Our model C data (SAW) comprise the largest range of chain lengths of our
three models, combined with precise estimates of statistical errors, which
allows a more detailed data analysis. For our $R_G$ data, we obtained the fit
$\left< R_G^2 \right> = A N^{1.1754} + B N^{0.62}$ with $A = 0.3341 \pm
0.0023$, $B = - 0.20 \pm 0.05$, where we again use the exponents from Ref.
\onlinecite{limadrassokal}. The deviation $\chi^2$ (sum of the residuals
squares, normalized by the variances) has the value $\chi^2 = 9.4$ (10 data
points). The corresponding quality of fit $Q$, which is the probability to
observe the measured $\chi^2$ value, or a larger one, is $Q = 0.31$. Our
data, in a representation which emphasizes the corrections to scaling, are
shown in Fig. \ref{fig:bdrg2corscal}. It is seen that these are indeed weak,
highlighting the difficulties in determining an accurate value for the
correction--to--scaling exponent.

Turning to our $R_H$ data from model C, we first did a nonlinear
two--parameter fit $\left< R_H^{-1} \right> = A N^{-\nu_{eff}}$, resulting in
$\nu_{eff} = 0.55$. However, this fit is very poor, with a least--square sum
$\chi^2 = 433$. Conversely, a linear two--parameter fit $\left< R_H^{-1}
\right> = A N^{-0.5877} + B N^{-1}$ yields a rather good value $\chi^2 = 11.8$
($Q = 0.16$), with $A = 2.732 \pm 0.005$, $B = -3.10 \pm 0.06$, demonstrating
also numerically that $R_H$ data should be interpreted in terms of corrections
to scaling, instead of an effective exponent. Actually, one should expect the
presence of an additional correction of order $N^{-\Delta}$, $\Delta \approx
0.56$. Since this correction tends to decrease $R_G$ (see Fig.
\ref{fig:bdrg2corscal}), it should also decrease $R_H$, i.~e. increase $\left<
  R_H^{-1} \right>$, or weaken the analytic $N^{-(1 - \nu)}$ 

\begin{figure}
\psfig{figure=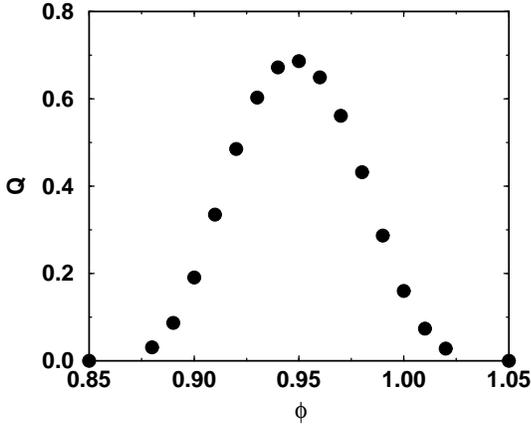,width=7cm}
\caption{Model C: Quality of the fit (see text) for the relation
  $\left< R_H^{-1} \right> = A N^{-0.5877} + B N^{- \phi}$ with fixed $\phi$,
  as a function of $\phi$.}
\label{fig:bdqvalues}
\end{figure}

\begin{figure}
\psfig{figure=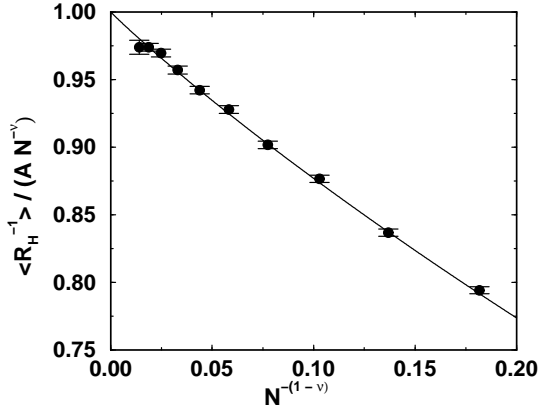,width=7cm}
\caption{Model C: $\left< R_H^{-1} \right> / (A N^{- \nu})$
  as a function of $N^{-(1 - \nu)}$, where we use $A = 2.753$ from the
  fit $\left< R_H^{-1} \right> = A N^{-0.5877} + B N^{-1} + C
  N^{-1.15}$, which is shown as well.}
\label{fig:bdrhcorscal}
\end{figure}

\begin{figure}
\psfig{figure=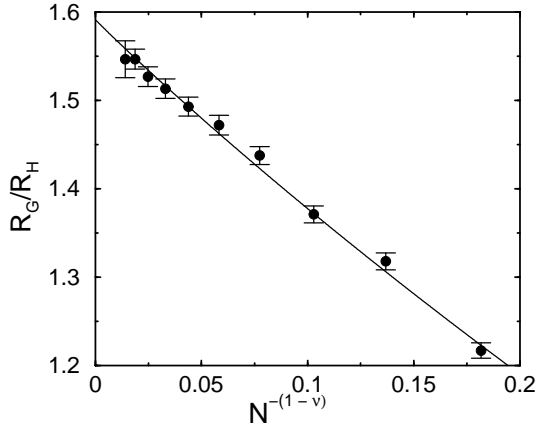,width=7cm}
\caption{Model C: $R_G / R_H$ as a function of the
   scaling variable $N^{-(1 - \nu)}$. The line results
   from the combined fits of $R_G$ and $R_H$.}
\label{fig:bdratio}
\end{figure}

term. Thus, in a
regression $\left< R_H^{-1} \right> = A N^{-0.5877} + B N^{-\phi}$, where we
keep $\phi$ fixed, one should obtain the best fit for a value of $\phi$
slightly smaller than unity. This is indeed what we observe, as seen from Fig.
\ref{fig:bdqvalues}, where we plot the quality $Q$ of such a fit as a function
of $\phi$. This figure also clearly rules out a single correction to scaling
with an exponent of $1/2$ or even larger. We thus attempted a three--parameter
fit $\left< R_H^{-1} \right> = A N^{-0.5877} + B N^{-1} + C N^{-1.15}$ to also
take the $N^{-\Delta}$ term into account. The result of this fit, which seems
to be reasonably stable, is $A = 2.753 \pm 0.008$, $B = - 4.3 \pm 0.4$, $C =
2.2 \pm 0.7$, with $\chi^2 = 5.0$, and a very good quality $Q = 0.66$. We thus
use this fit to demonstrate the corrections to scaling of $\left< R_H^{-1}
\right>$ in Fig. \ref{fig:bdrhcorscal}, where the presence of the
$N^{-\Delta}$ term shows up in a slight curvature. Finally, we also used this
fit, combined with the corresponding one for $R_G$ (see Fig.
\ref{fig:bdrg2corscal}), to describe the data on the ratio $\rho = R_G / R_H$,
as shown in Fig. \ref{fig:bdratio}, where the asymptotic value is $1.591 \pm
0.007$. Again we feel that the real uncertainty is larger, due to lack of
control of the systematic errors. We also checked that both the quality of
fit, and the value of $\rho$ did not change significantly when we reduced the
exponent $\Delta$ to its theoretical value \cite{zinnjustin} $\Delta = 0.482$.

To summarize, we have collected our most important numerical results,
the extrapolated $\rho$ values, in Table \ref{tab:rgrhvalues}.

\begin{table}
  \begin{center}
    \begin{tabular}{| c | c |} \hline
    Model & $R_G / R_H$ \\
    \hline \hline
    A (SAW)      & $1.63  \pm 0.01$  \\
    B (SAW)      & $1.61  \pm 0.02$  \\
    C (SAW)      & $1.591 \pm 0.007$ \\
    A ($\Theta$) & $1.44  \pm 0.01$  \\
    \hline
    \end{tabular}
    \caption{Asymptotic universal ratio $R_G / R_H$ as estimated
             by numerical simulations of various models (see text).
             Error bars take into account statistical
             uncertainties only, while systematic errors in the
             extrapolation procedure are neglected.}
    \label{tab:rgrhvalues}      
  \end{center}
\end{table}

\section*{Acknowledgments}

Stimulating discussions with J. Horbach, A. J. C. Ladd, J. J. de Pablo, and S.
Wiegand are gratefully acknowledged. We thank L. Sch\"afer, A. Z. Akcasu and
Y. Tsunashima for useful remarks and hints to the literature, G. Besold for a
critical reading of the manuscript, and DSM and the BMBF Competence Center in
Materials Simulations for financial support.

\begin{appendix}
\section{Euler--Maclaurin Formula}
\label{sec:euler}

Quite usually, sums are approximated via the corresponding integrals. The
Euler--Maclaurin formula \cite{olver,graham}, which we outline here for the
convenience of the reader, constructs a systematic asymptotic expansion around
that approximation. Defining a difference operator $\Delta$ via
\begin{equation}
\Delta f(x) = f(x + 1) - f(x),
\end{equation}
one obviously has
\begin{equation}
\Delta F(N) = f(N)
\end{equation}
for
\begin{equation}
F(N) = \sum_{n = n_0}^{N - 1} f(n) ,
\end{equation}
and thus
\begin{equation}
F(N) = \Delta^{-1} f(N) + \mbox{\rm const.} .
\end{equation}
On the other hand,
\begin{equation}
\Delta = \exp \left( \frac{d}{dx} \right) - 1
\end{equation}
or
\begin{eqnarray}
\Delta^{-1} & = & \left( \frac{d}{dx} \right)^{-1}
\left( \frac{d}{dx} \right)
\left[ \exp \left( \frac{d}{dx} \right) - 1 \right]^{-1}
                 \nonumber \\
& = & \int dx
\sum_{k = 0}^\infty \frac{B_k}{k !}  \left( \frac{d}{dx} \right)^k ,
\end{eqnarray}
where $B_k$ are the Bernoulli numbers defined via the Taylor expansion of $x /
(e^x - 1)$: $B_0 = 1$, $B_1 = -1/2$, $B_2 = 1/6$, $B_4 = -1/30$, \ldots, $B_3
= B_5 = B_7 = \ldots = 0$. Hence,
\begin{equation}
\Delta^{-1} = \int dx - \frac{1}{2} + \frac{1}{12} \frac{d}{dx}
- \frac{1}{720} \left( \frac{d}{dx} \right)^3 + \ldots
\end{equation}
and thus
\begin{eqnarray} \label{eq:eulermaclaurin}
\sum_{n = n_0}^{N - 1} f(n) & = & \int_{n_0}^N dx f(x) - \frac{1}{2} f(N)
+ \mbox{\rm const.}  \nonumber \\
& + & \frac{1}{12} \left. \frac{d}{dx} f(x) \right\vert_{x = N}
- \frac{1}{720} \left. \frac{d^3}{dx^3} f(x) \right\vert_{x = N}
\nonumber \\ & + & \ldots ,
\end{eqnarray}
where the ``integration'' constant is determined via (perhaps numerical)
comparison of both sides. For a power law with $q < -1$ one thus finds from
the definition of the Riemann zeta function
\begin{eqnarray} \label{eq:eulermaclaurinpower}
\sum_{n = 1}^{N - 1} n^q & = & \frac{N^{q + 1}}{q + 1} - \frac{1}{2} N^q
+ \zeta(-q) + \frac{1}{12} q N^{q - 1} \nonumber \\
& - & \frac{1}{720} q (q - 1) (q - 2) N^{q - 3} + \ldots .
\end{eqnarray}
By analytic continuation with respect to $q$, this result holds for general
$q$ \cite{olver}.

\end{appendix}

\end{multicols}


\begin{thebibliography}{10}

\bibitem{limadrassokal}
B. Li, N. Madras, and A.~D. Sokal, J. Stat. Phys. {\bf 80},  661  (1995).

\bibitem{doiedw}
M. Doi and S.~F. Edwards, {\em The Theory of Polymer Dynamics} (Clarendon
  Press, Oxford, 1986).

\bibitem{sokal}
A.~D. Sokal,  in {\em Monte Carlo and Molecular Dynamics Simulations in Polymer
  Science}, edited by K. Binder (Oxford University Press, New York, 1995), p.\
  47.

\bibitem{besold}
G. Besold, H. Guo, and M.~J. Zuckermann, J. Polym. Sci., Polym. Phys. {\bf 38},
   1053  (2000).

\bibitem{schaefer}
L. Sch{\"a}fer, {\em Excluded Volume Effects in Polymer Solutions as Explained
  by the Renormalization Group} (Springer, Berlin, 1999).

\bibitem{zinnjustin}
J. Zinn-Justin, {\em Quantum Field Theory and Critical Phenomena}, 2nd ed.
  (Clarendon Press, Oxford, 1993).

\bibitem{adamdelsanti1}
M. Adam and M. Delsanti, J. Physique {\bf 37},  1045  (1976).

\bibitem{adamdelsanti2}
M. Adam and M. Delsanti, Macromolecules {\bf 10},  1229  (1977).

\bibitem{nemoto}
N. Nemoto, Y. Makita, Y. Tsunashima, and M. Kurata, Macromolecules {\bf 17},
  425  (1984).

\bibitem{venkataswamy}
K. Venkataswamy, A.~M. Jamieson, and R.~G. Petschek, Macromolecules {\bf 19},
  124  (1986).

\bibitem{simone}
D. Reith, B. M{\"u}ller, F. M{\"u}ller-Plathe, and S. Wiegand, preprint,
  submitted to Macromolecules, cond-mat/0103329.

\bibitem{tsunashima1}
Y. Tsunashima, M. Hirata, N. Nemoto, and M. Kurata, Macromolecules {\bf 20},
  1992  (1987).

\bibitem{depablo}
R.~M. Jendrejack, M.~D. Graham, and J.~J. de~Pablo, J. Chem. Phys. {\bf 113},
  2894  (2000).

\bibitem{weilldescloizeaux}
G. Weill and J. des Cloizeaux, J. Physique {\bf 40},  99  (1979).

\bibitem{schaeferbaumgaertner}
L. Sch\"afer and A. Baumg\"artner, J. Physique {\bf 47},  1431  (1986).

\bibitem{kkdipl}
K. Kremer, A. Baumg{\"a}rtner, and K. Binder, Z. Phys. B {\bf 40},  33  (1981).

\bibitem{oonokohmoto}
Y. Oono and M. Kohmoto, J. Chem. Phys. {\bf 78},  520  (1983).

\bibitem{tsunanew}
Y. Tsunashima, Macromolecules {\bf 21},  2575  (1988).

\bibitem{douglasfreed}
J.~F. Douglas and K.~F. Freed, Macromolecules {\bf 17},  2354  (1984).

\bibitem{guttman1}
C.~M. Guttman, F.~L. McCrackin, and C.~C. Han, Macromolecules {\bf 15},  1205
  (1982).

\bibitem{guttman2}
C.~M. Guttman, J. Stat. Phys. {\bf 36},  717  (1984).

\bibitem{akcasnew}
A.~Z. Akcasu and C.~M. Guttman, Macromolecules {\bf 18},  938  (1985).

\bibitem{hager}
J. Hager and L. Sch\"afer, Phys. Rev. E {\bf 60},  2071  (1999).

\bibitem{grassberger}
P. Grassberger and R. Hegger, J. Chem. Phys. {\bf 102},  6881  (1995).

\bibitem{schmidtburchard}
M. Schmidt and W. Burchard, Macromolecules {\bf 14},  211  (1981).

\bibitem{tsunashima2}
Y. Tsunashima {\it et~al.}, Macromolecules {\bf 20},  2862  (1987).

\bibitem{batouliskremer}
J. Batoulis and K. Kremer, Macromolecules {\bf 22},  4277  (1989).

\bibitem{laddfrenkel}
A.~J.~C. Ladd and D. Frenkel, Macromolecules {\bf 25},  3435  (1992).

\bibitem{akcasugurol}
A.~Z. Akcasu and H. Gurol, J. Polym. Sci., Polym. Phys. {\bf 14},  1  (1976).

\bibitem{benmounaakcasu}
M. Benmouna and A.~Z. Akcasu, Macromolecules {\bf 13},  409  (1980).

\bibitem{hanakcasu}
C.~C. Han and A.~Z. Akcasu, Macromolecules {\bf 14},  1080  (1981).

\bibitem{bhatt1}
M. Bhatt and A.~M. Jamieson, Macromolecules {\bf 21},  3015  (1988).

\bibitem{bhatt2}
M. Bhatt, A.~M. Jamieson, and R.~G. Petschek, Macromolecules {\bf 22},  1374
  (1989).

\bibitem{shiwa}
Y. Shiwa, J. Phys. A, Math. Gen. {\bf 24},  L579  (1991).

\bibitem{tsunashima3}
Y. Tsunashima, Polym. Journ. {\bf 24},  433  (1992).

\bibitem{bdphd}
B. D{\"u}nweg and K. Kremer, J. Chem. Phys. {\bf 99},  6983  (1993).

\bibitem{hammouda}
B. Hammouda and A.~Z. Akcasu, Macromolecules {\bf 16},  1852  (1983).

\bibitem{tsunaprivate}
Y. Tsunashima, private communication.

\bibitem{kremer}
K. Kremer,  in {\em Monte Carlo and Molecular Dynamics of Condensed Matter
  Systems}, edited by K. Binder and G. Ciccotti (Italian Physical Society,
  Bologna, 1996), p.\ 669.

\bibitem{thoso}
T. Soddemann, B. D{\"u}nweg, and K. Kremer, Europ. Phys. J. E {\bf 6},  409
  (2001).

\bibitem{biermann01s}
O. Biermann {\it et~al.}, preprint, submitted to Macromolecules,
  cond-mat/0101115.

\bibitem{reith00s}
D. Reith, H. Meyer, and F. M{\"u}ller-Plathe, Macromolecules {\bf 34},  2335
  (2001).

\bibitem{olver}
F.~W.~J. Olver, {\em Asymptotics and Special Functions} (Academic Press, New
  York, 1974).

\bibitem{graham}
R.~L. Graham, D.~E. Knuth, and O. Patashnik, {\em Concrete Mathematics}
  (Addison--Wesley, Boston, 1994).

\end{thebibliography}
\end{document}